\documentclass[12pt, reqno]{amsart}
\usepackage{amssymb,amsmath,mathrsfs,MnSymbol,setspace, xcolor}
\usepackage{setspace}
\usepackage{graphicx}
\usepackage{epsfig}
\makeatletter \oddsidemargin.9375in \evensidemargin \oddsidemargin
\def\authorsaddresses#1{\dedicatory{#1}}

\newtheorem{theorem}{theorem}[section]

\theoremstyle{definition}

\theoremstyle{remark}
\newtheorem{remark}[theorem]{Remark}
\numberwithin{equation}{section}

\begin{document}
\setcounter{page}{1}


\title[An Alternating Qubit Protocol and Its Correctness Checking]{ An Alternating Qubit Protocol and Its Correctness Checking}

\author[H. Farahani]{H. Farahani}


\authorsaddresses{Department of Computer Science, Shahid Beheshti University, G.C, Tehran, Iran\\
h$_-$farahani@sbu.ac.ir}
%
%
\begin{abstract}
In this paper, a quantum version of classical alternating bit protocol is proposed. This protocol provides a reliable method to transmit the secret quantum data via a noisy quantum channel while the entanglement between particles is not broken. Our protocol is based on quantum teleportation and superdense coding. By assuming that the participants can distinguish the alternating qubit from other messages and also the assumption that data can be resent unlimited times, an abstraction of this protocol can be derived. Using the quantum process algebra \textit{full} $qACP$, we show that the proposed protocol is correct, so the desired external behaviour of the protocol is guaranteed.
\end{abstract}
\maketitle
\section{Introduction}
%
%
%
Alternating bit protocol was proposed to achieve reliable full-duplex operation over half-duplex classical lines  \cite{Alternatingbit}. In this protocol each transmitted message contains error detection information and a control bit is used as a "validation" to indicate correct or incorrect 
 arrival of a message. The validation bit alternates such that a change in its value means "acknowledgement". This bit is called the \textit{alternating bit}.
%
%
$ $\\
In quantum communications, noisy quantum channels are subject to information corruption and loss. There are different types of noise and noisy channels such as bit-flip, phase-flip, erasure  etc. \cite{Wilde2013}. These noises may change the state of communicating quantum particles.

Analogous to classical counterparts, a theory of quantum error-correction has been proposed which allows effective computations in the presence of noise and  reliable communication over noisy quantum channels. Quantum error-correcting codes combat the effects of noise and allow reliable communication and computation even in the presence of quite severe noise \cite{Nielsen}.
The error syndrome can be detected by error syndrome measurement which does not cause any change to the state of quantum particle and indicates the type of  occurred error. After the measurement, an appropriate procedure can be used to recover the initial state exactly or with high probability \cite{Nielsen}.

In Section 2, a \textit{quantum alternating qubit protocol} (briefly, QAQP)  as a quantum version of alternating bit protocol is proposed. This protocol is based on quantum teleportation and superdense coding \cite{Nielsen} and introduces a method to ensure reliable transmission of quantum data through a noisy quantum channel. 
It is assumed that data can be resent unlimited times and the noisy channel does not break the entanglement between particles.

The process algebra ACP (Algebra of Communicating Processes) is one of the standard approaches of correctness checking \cite{Fokkink}. qACP is a quantum version of  ACP which applies a special quantum process configuration to use both quantum  and classical information (computation).
In \cite{Wang2013} a relationship between quantum and classical bisimularity is obtained in a way that qACP and classical ACP are unified. This unification is important in verification of quantum communication protocols, since most of these protocols including QAQP involve both classical and quantum information and/or computation. 

As it is mentioned above, QAQP is based on quantum teleportation and superdense coding, so the entangled particles are used and we need a quantum process algebra which consider the entanglement. Hence, we use the \textit{full qACP} instead of qACP to model the entanglement as a kind of parallelization for checking the correcness of QAQP.  The \textit{full qACP} is an extension of qACP equipped with a \textit{shadow constant} and an \textit{entanglement merge} \cite{Wang2015}. This extension has a sound and complete axiomatization modulo a quantum bisimilarity and is strong enough to verify any quantum protocol  that uses both classical and quantum information and also adopt entanglement in some steps.

Section 3 is devoted to specification of QAQP by \textit{ full }$qACP$. A linear recursive specification for participants is given by the assumption that the recovery procedures always recover the initial state and  that the difference between the initial state before recovery and the outcome state after recovery  is negligible. Furthermore,  the encoding, syndrome measurement and recovery processes are not considered in our abstraction. Note that the correctness of QAQP can also be checked by the process algebra qCCS \cite{Feng2014}.

In Section 4, by identifying communication actions, encapsulation and abstraction operations, the formal verification of the QAQP is given to show that this protocol has the desired external behaviour.
\section{Quantum alternating qubit protocol}

We assume that the reader is familiar with the quantum teleportation and superdense coding. This protocol has two participants sender (Alice) and receiver (Bob), which Alice receives some qubits from an isolated private quantum channel Q. Alice must deliver them to Bob and then Bob has to send them into an isolated private quantum channel P. 

The secret quantum data $q_1, q_2,...,q_r$ from a finite set $\Delta$ are communicated between Alice and Bob.
Every communication between Alice and Bob is performed via the only noisy quantum channel D such that Alice is not permitted to use this quantum channel to communicate the secret quantum data. 
In the absence of a legal quantum communicating channel, quantum teleportation is used. An EPR pair $(A,B)$ is generated, then Alice and Bob take 
one of the EPR pair's qubits before starting the protocol. 
Alice interacts the secret quantum data $q_i$ with her half of EPR pair $A$, then performs a measurement and obtains
one of the four possible classical results 00, 01, 10, and 11. She must send two classical outcome bits to Bob. 

In order to communicate these classical bits, the superdense coding is applied. However, since the channel D is corrupted, an \textit{alternating qubit} is communicated between them to ensure Alice that Bob has received the desired classical bits. In other word, this alternating qubit is an\textit{ acknowledgement} from Bob.
%
%
If Alice reads the secret quantum datum $q_i$ from channel Q with an \textit{odd} index, 
she sends the qubit $b=|0\rangle$ to Bob through quantum channel D and must receive the same qubit as acknowledgement from Bob to start the communication. Likewise, the qubit $b=|1\rangle$ for even indices. 

Since the channel D is noisy, it is possible that the communicated message through this channel has turned into an \textit{error message}, which is shown by $\perp$. 
The error syndrome can be detected by an error syndrome measurement which although does not cause any change to the state of quantum particle, but indicates what kind of error has occurred. After the measurement, an appropriate procedure can be used to recover the initial state exactly or with high probability.


Now, we explain the protocol in more details. In the sequel, we shall use the following conventions: by "sending or receiving a qubit", we mean "sending or receiving an \textit{encoded} qubit" and for simplicity we show the resulting encoded qubit $S$ by $S$ itself. Also, by \textit{correct} qubit we mean that after performing an appropriate error syndrome measurement, no error has been detected.
\begin{figure} [htbp]
\vspace*{13pt}
\centerline{\epsfig{file=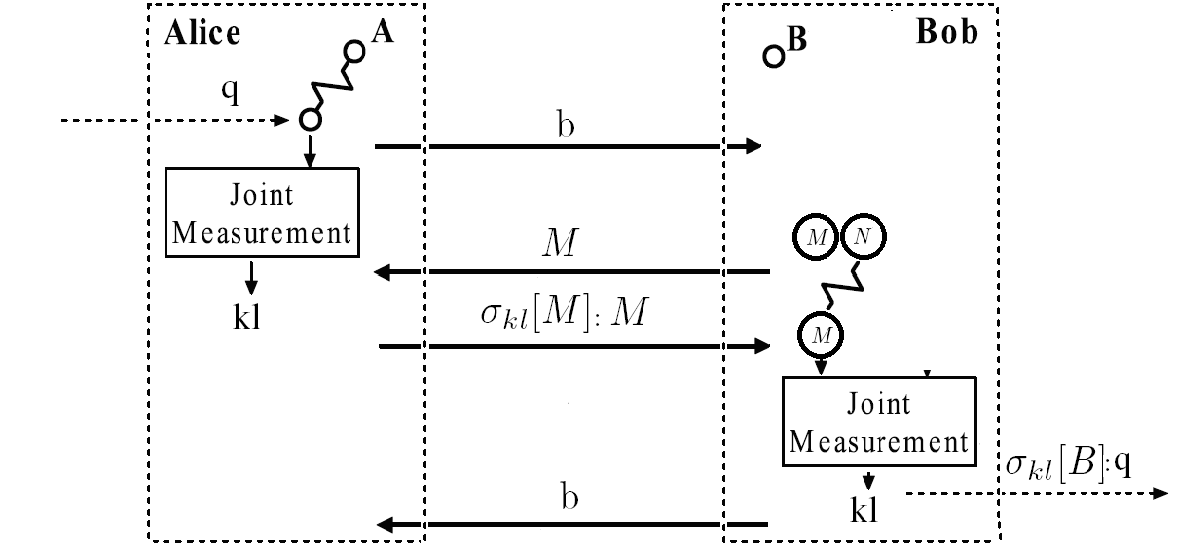, width=8.2cm}} 
\vspace*{13pt}
\caption{Quantum Alternating Qubit Protocol.}
\end{figure}


As shown in Fig. 1, QAQP is started by sending an encoded qubit $b$ from Alice to Bob into the noisy quantum channel D. If Bob receives the correct qubit $b$, then he prepares an EPR pair ($M,N$) and 
sends the qubit M to Alice through channel D. Otherwise, he sends back the error message $\perp$ to Alice and waits to receive the qubit $b$ from Alice. Note that in this case when Alice receives the error message, she will send the qubit $b$ to Bob again. 
%
 Since the quantum channel D is noisy, in the case that Bob sends the qubit M, some errors may be occurred on it. 
 If Alice detects any error, she will send the qubit $b$ to Bob again. 
 Otherwise, she makes a joint measurement on the secret quantum data $q_i$ with her half $A$ of EPR pair of quantum teleportation which has been fixed between Alice and Bob before starting the protocol. Corresponding to the classical result $kl\in\{00, 01, 10, 11\}$ of the measurement, Alice performs the desired Pauli's operator $\sigma_{kl}$ on the qubit $M$. Then, she encodes the outcome and sends it to Bob through quantum channel D and waits to receive the acknowledgement from Bob.

 Because of the noisy channel D, the sent message can be turned into an error. Therefore, Bob performs an error syndrome measurement. If Bob detects an error, 
 he generates a fresh EPR pair $[M,N]$ and sends $M$ to Alice. Otherwise, Bob sends the acknowledgement qubit $b$ to Alice and then decodes the received message M. 
 In the sequel, he performs a joint measurement on the qubits $M$ and $N$ that he has in hand. For the convenient, by $M$ we denote the resulting qubit after encoding and/or performing the Pauli's operator $\sigma_{kl}$ on the qubit $M$ and also the fresh EPR pair is shown by $[M,N]$ again.

Afterwards, Bob registers the measurement result $kl$ and performs the corresponding operator $\sigma_{kl}$ on the photon B which was fixed between participants before starting the protocol in order to use the quantum teleportation. Finally, Bob sends the outcome qubit into quantum channel P and goes to new state to be ready for new session of the protocol.
If Alice receives the acknowledgement qubit $b$ successfully, she sends the qubit $1-b$ to start a new session. 
Otherwise, she sends qubit $b$ to Bob to inform him that the received acknowledgement is corrupted and Bob has to send the acknowledgement again. Note that for simplicity we use $1-b$. If $b=|0\rangle$ then $1-b=|1\rangle$ else $b=|1\rangle$ then $1-b=|0\rangle$.

Moreover, in QAQP, we suppose that participants can distinguish the acknowledgement qubit from other message. We call this assumption the \textit{distinguishing assumption}, which allows us  to define an easier form of abstraction.


\section{Specification of QAQP by full $qACP$}

%

In this section a linear recursive specification of the QAQP is given by process algebra \textit{full }$qACP$. We assume that the recovery procedures always recover the initial state. Also the encoding, syndrome measurement and recovery steps are not considered in the abstraction. 

We prove that the resulting algebraic process term displays the desired external behaviour of the protocol, that is the secret quantum data read from channel Q by Alice are sent into channel P by Bob in the same order without losing any data element. The process term is a solution for the following recursive specification:
$$X=\sum\limits_{q_i\in \Delta} read_{Q}[q_i]. send_{P}[q_i].X$$
where, the action $read_{Q}[q_i]$ represents "\textit{read datum $q_i$ from channel Q}" and the action
$send_P[q_i]$ represents "\textit{send datum $q_i$ into channel P}".

First, we specify Alice in the state that she is going to send out data with the acknowledgement qubit $b$, represented by the recursion variable $S(b)$ for $b\in\{|0\rangle,|1\rangle\}$:
\\\\
$S(b)=\sum\limits_{q_i \in \Delta } read_Q[q_i].S_1$\\
$S_1= (send_D[b] +send_D[\perp]). S_2$\\
$S_2=receive_D[M]. S_3 +receive_D[\perp].S_1+ receive_D[b].S(1-b)$\\
$S_3= Me[A,q_i; kl]. S_4$\\
$S_4=\sigma_{kl}[M].S_5$
\\$S_5=(send_D[M]+send_D[\perp]).S_6$\\
$S_6=receive_D[b].S(1-b)+receive_D[M].S_4+receive[\perp].S_1$\\

In state S(b), Alice reads the quantum datum $q_i$ from channel Q. Then the system proceeds to
state $S_1$, in which Alice sends the qubit $b$ into channel D. However, the qubit $b$ may be distorted by the noisy channel, so that it may proceeds to an error message $\perp$. Next the system goes to state $S_2$ and Alice waits to receive the qubit $M$ through channel D, but the message $M$ may be turned into an error. So, after receiving the message, Alice performs an error syndrome measurement to detect the conceivable error $\perp$. 
If no error has been detected, she recovers the qubit $M$ and the system proceeds to state $S_3$. If she detects any error, the system traverse to state $S_1$ again.
In state $S_2$, there is also another case, which Alice receives the qubit $1-b$ or the noisy $1-b$. By our distinguishing assumption, Alice can distinguish this message from messages $M$ or noisy $M$. This case occurs when some secret data $q_i$ has communicated between Alice and Bob previously and the last correct acknowledgement from Bob has not received by Alice. 
In this case, Bob had resent the acknowledgement of Alice's previous request and after receiving the correct acknowledgement by Alice, the system proceeds to $S(1-b)$ to start new communication session.

In state $S_3$, Alice performs a joint measurement on qubits $A$ and $q_i$ and registers the result as $kl$. Then, the system proceeds to state $S_4$ and Alice operates $\sigma_{kl}$ on $M$.
Next, in state $S_5$, Alice sends the outcome $M$ to Bob into channel D. However, since the channel D is noisy the message may be turned into an error.
The next state of the system is $S_6$. If Alice receives the correct acknowledgement $b$, the system proceeds to state $S(1-b)$ to start new session. Otherwise, there are two cases; either Alice receives qubit $M$ or Alice detects some errors. The system goes to states $S_4$ or $S_1$, respectively. 

In the case that Alice receives the qubit $M$, it is perceived that in the last step of the protocol, Bob has received the corrupted message $M$ and so he has generated a new EPR pair $[M',N']$ and has sent the qubit $M'$ to Alice. 
In order to prevent ambiguity in abstraction of the protocol, we show the fresh qubits $M'$ and $N'$ by $M$ and $N$ again.

Next, we specify Bob in the state that he is expecting to receive the qubit $b$, represented by the recursion variable $R(b)$, for $b\in\{|0\rangle,|1\rangle\}$:
\\\\
$R(b)=receive_D[b]. R_2 + ( receive_D[\perp]+receive_D[1-b]).R_1 $\\
$R_1 = (send_D[\perp]+send_D[1-b]).R(b)$\\
$R_2 = GEN[M,N].R_3$\\
$R_3 = (send_D[M]+send_D[\perp]).R_4$\\
$R_4=\circledS_{Me[A,q_i; kl]}.R_5+\circledS_{\sigma_{kl}[M]}.R_6+ (receive_D[b] + receive_D[\perp]).R_2$\\
$R_5=\circledS_{\sigma_{kl}[M]}.R_6$\\
$R_6 = receive_D[M].R_7 + receive_D[\perp].R_2$\\
$R_7=(send_D[b]+send_D[\perp]).R_8$\\
$R_8=Me[M,N; kl].R_9$\\
$R_9=\sigma_{kl}[B].R_{10}$\\
$R_{10}=send_P[B].R(1-b)$.\\
\\
In state $R(B)$, if Bob receives the correct alternating qubit $b$, the system proceeds to $R_2$. Otherwise, there are two cases; either Bob reads an error message $\perp$ from the channel D, that this does not constitute new information and he sends $\perp$ back into the channel D or Bob receives $1-b$ which indicates that in the previous session of the protocol, Alice has not received the acknowledgement successfully. In this case Bob sends $1-b$ to Alice again and the system goes to state $R(b)$.

In state $R_2$, Bob generates an EPR pair $(M, N)$, then the system proceeds to $R_3$ and he sends the qubit $M$ to Alice through the channel D. However, the channel D is noisy and 
the message may  turn into an error. 
Afterwards, Bob awaits to receive the message $M$, but if he receives the qubit $b$ or the corrupted qubit $b$ (in state $R_4$), by distinguishing assumption it can be perceived that the sent qubit $M$ from Bob to Alice in state $R_3$ is corrupted and Alice in state $S_2$ has received the noisy message and she has tried to send the qubit $b$ to announce Bob for the corruption. 
In state $R_6$, if he receives the correct message $M$, the system goes to state $R_7$ and he sends the acknowledgement $b$ to Alice. Otherwise the system proceeds to state $R_2$ again. 
In the next state, Bob makes a joint measurement on particles $M$ and $N$, and registers the result as $kl$. In state $R_9$, the operator $\sigma_{kl}$ is applied on the qubit $B$. 
Finally, in state $R_{10}$, Bob sends the outcome of the state $R_9$ into channel P and the state of the system goes to R(1-b) to start a new communication session.
\\

\section{Verification of QAQP}
In the sequel, communication actions, encapsulation and abstraction operations are explained. Send/receive actions of the same message (the qubits $b$ or $M$ or $\perp$) over the channel D, communicate with each other. Thus for each qubit $M$ and every $b\in\{|0\rangle,|1\rangle\}$ we have the following communications:
$$\gamma(send_D[b], receive_D[b]) = C_D[b]$$
$$\gamma(send_D[\perp], receive_D[\perp]) = C_D[\perp]$$
$$\gamma(send_D[M], receive_D[M]) = C_D[M]$$
$$\gamma(send_D[1-b], receive_D[1-b]) = C_D[1-b]$$

 All other communications between atomic actions result in $\delta$. Note that atomic actions in QAQP are: 
 receiving quantum data through channel Q, sending (receiving) quantum data or $\perp$ into (through) channel D, sending data into channel P, generating an EPR pair, performing a joint measurement on two qubits and registering the outcome, applying a Pauli's operator on a particle. Some another actions such as encoding, syndrome measurement and recovery are not considered  in the abstraction of QAQP.

The desired system is obtained by putting $R(0)$ and $S(0)$ in parallel, encapsulating some 
actions over the quantum channel D and abstracting away from communication actions over this channel. Therefore, QAQP is expressed by the following process term: $$\tau_I(\partial_H(S(0)||R(0))$$
Where, 
$x||y=(x|\lfloor y+y|\lfloor x)+x|y+x\between y$. $H$ and $I$ are defined as follows:
\begin{align*}
H=\{&send_D[b], receive_D[b], send_D[1-b], receive_D[1-b], send_D[\perp], receive_D[\perp],\\
& send_D[M], receive_D[M], Me[A,q_i; kl], \circledS_{Me[A,q_i; kl]}, \sigma_{kl}[M], \circledS_{\sigma_{kl}[M]}\}
\end{align*}
%
%
$I= \{C_D[b], C_D[M], C_D[\perp], \sigma_{kl}[M], GE[M,N], Me[A,q_i; kl], Me[M,N;kl], \sigma_{kl}[B]\}$
\\
\\
where $M, N, A, B, q_i$ are qubits and $b\in\{|0\rangle,|1\rangle\}$.
\\

Now, in order to proceed the 
 formal verification of QAQP, some axioms of full ${{qACP}}$ are restated from \cite{Wang2015}:
$$\begin{array}{|l|l|}
\hline 
 Name& ~~~~~~~~~~~~~~~~~~~~~~~~~~~~~~~~~~Axiom \\ 
 \hline 
 QTI1 & v\notin I~~\tau _I(v)=v \\ 
QTI2 & v \in I ~~\tau _I(v)=\tau \\ 
QTI3 & \tau _I(\delta)=\delta \\ 
QTI4 & \tau _I(x+y)=\tau _I(x)+\tau _I(y) \\ 
QTI5 & \tau _I(x.y)=\tau _I(x).\tau _I(y) \\ 
RDP & \langle X_i | E\rangle = t_i(\langle X_1 |E\rangle ,. . . ,\langle X_n |E\rangle),~\text{for} ~ i\in \{ 1,2,...,n\}\\ 
RSP & \text{If}~ y_i = t_i(y_1, ...,y_n)~ \text{then}~  y_i = \langle X_i | E \rangle, \text{for} ~ i\in \{ 1,2,...,n\} \\ 
CFAR & \text{If X is in a cluster for I with exits}~ \{v_1Y_1,...,v_mY_m, w_1,...w_n\}, \text{then}\\ 
& \tau.\tau_I(\langle X|E \rangle)=\tau.\tau_I(v_1\langle Y_1|E \rangle,...v_m\langle Y_m|E \rangle,w_1,...,w_n)\\
 \hline 
\end{array}$$

  $ $\\ \\The following equations are derived from 
 the axioms of full ${qACP}$ and RDP:
\\
\\
$\partial_H(S(0)||R(0)) =\sum\limits_{q_i \in \Delta } read_Q[q_i] .\partial_H(S_1||R(0))$\\
$\partial_H(S_1||R(0)) =C_D[0] .\partial_H(S_2||R_2) + C_D[\perp].\partial_H(S_2||R_1)$\\
$\partial_H(S_2||R_1) =C_D[\perp].\partial_H(S_1||R(0))$\\
$\partial_H(S_2||R_2) =GEN[M,N].\partial_H(S_2||R_3)$\\
$\partial_H(S_2||R_3)=C_D[M].\partial_H(S_3||R_4)+C_D[\perp].\partial_H(S_1||R_4) $\\
$\partial_H(S_1||R_4)=(C_D[0]+C_D[\perp]).\partial_H(S_2||R_2)$\\ 
$\partial_H(S_3||R_4) =Me[A,q_i; kl].\partial_H(S_4||R_5)$\\
$\partial_H(S_4||R_5) =\sigma_{kl}[M] .\partial_H(S_5||R_6)$\\
$\partial_H(S_5||R_6)=C_D[M].\partial_H(S_6||R_7)+C_D[\perp].\partial_H(S_6||R_2) $\\
$\partial_H(S_6||R_2) =GEN[M,N].\partial_H(S_6||R_3)$\\
$\partial_H(S_6||R_3)=C_D[M].\partial_H(S_4||R_4)+C_D[\perp].\partial_H(S_1||R_4) $\\
$\partial_H(S_4||R_4) =\sigma_{kl}[M]\partial_H(S_5||R_6)$\\
$\partial_H(S_6||R_7) =C_D[0].\partial_H(S(1)||R_8)+C_D[\perp] .\partial_H(S_1||R_8)$\\
$\partial_H(S_1||R_8) =Me[M,N; kl].\partial_H(S_1||R_9)$\\
$\partial_H(S_1||R_9) =\sigma_{kl}[B].\partial_H(S_1||R_{10})$\\
$\partial_H(S_1||R_{10}) =send_P[B].\partial_H(S_1||R(1))$\\
$\partial_H(S_1||R(1)) =(C_D[0]+C_D[\perp]).\partial_H(S_2||R_1)$\\
$\partial_H(S(1)||R_8) =Me[M,N;kl].\partial_H(S(1)||R_9)$\\
$\partial_H(S(1)||R_9) =\sigma_{kl}[B].\partial_H(S(1)||R_{10})$\\
$\partial_H(S(1)||R_{10}) =send_P[B].\partial_H(S(1)||R(1))$\\
\\
The process term $S(0)||R(0)$ can be expanded and each equation can be easily proved.
Note that the process term $\partial_H(S(1)||R(1))$ in the right-hand side of the above last equation is
not as the left-hand side of any equation. Below it proceeds to expand $\partial_H(S(1)||R(1))$. 
That is similar to above equations.
\\
\\
$\partial_H(S(1)||R(1)) =\sum\limits_{q_i \in \Delta } read_Q[q_i] .\partial_H(S_1||R(1))$\\
$\partial_H(S_1||R(1)) =C_D[1] .\partial_H(S_2||R_2) + C_D[\perp].\partial_H(S_2||R_1)$\\
$\partial_H(S_2||R_1) =C_D[\perp].\partial_H(S_1||R(1))$\\
$\partial_H(S_2||R_2) =GEN[M,N].\partial_H(S_2||R_3)$\\
$\partial_H(S_2||R_3)=C_D[M].\partial_H(S_3||R_4)+C_D[\perp].\partial_H(S_1||R_4) $\\
$\partial_H(S_1||R_4)=(C_D[1]+C_D[\perp]).\partial_H(S_2||R_2)$\\ 
$\partial_H(S_3||R_4) =Me[A,q_i; kl].\partial_H(S_4||R_5)$\\
$\partial_H(S_4||R_5) =\sigma_{kl}[M] .\partial_H(S_5||R_6)$\\
$\partial_H(S_5||R_6)=C_D[M].\partial_H(S_6||R_7)+C_D[\perp].\partial_H(S_6||R_2) $\\
$\partial_H(S_6||R_2) =GEN[M,N].\partial_H(S_6||R_3)$\\
$\partial_H(S_6||R_3)=C_D[M].\partial_H(S_4||R_4)+C_D[\perp].\partial_H(S_1||R_4) $\\
$\partial_H(S_4||R_4) =\sigma_{kl}[M]\partial_H(S_5||R_6)$\\
$\partial_H(S_6||R_7) =C_D[1].\partial_H(S(0)||R_8)+C_D[\perp] .\partial_H(S_1||R_8)$\\
$\partial_H(S_1||R_8) =Me[M,N; kl].\partial_H(S_1||R_9)$\\
$\partial_H(S_1||R_9) =\sigma_{kl}[B].\partial_H(S_1||R_{10})$\\
$\partial_H(S_1||R_{10}) =send_P[B].\partial_H(S_1||R(1))$\\
$\partial_H(S_1||R(0)) =(C_D[1]+C_D[\perp]).\partial_H(S_2||R_1)$\\
$\partial_H(S(0)||R_8) =Me[M,N;kl].\partial_H(S(0)||R_9)$\\
$\partial_H(S(0)||R_9) =\sigma_{kl}[B].\partial_H(S(0)||R_{10})$\\
$\partial_H(S(0)||R_{10}) =send_P[B].\partial_H(S(0)||R(0))$\\
\\
Owing to all above equations, RSP yields
$$\partial_H(R(0)||S(0))=\langle X_1|E \rangle~~~~~~~~~~(*)$$
where E denotes the following linear recursive specification
$$\begin{array}{|l|l|}
\hline
 & \\
X_1 =\sum\limits_{q_i \in \Delta } read_Q[q_i] .X_2 & Y_1 =\sum\limits_{q_i \in \Delta } read_Q[q_i] .Y_2\\
X_2 =C_D[0] .X_3 + C_D[\perp].X_4 & Y_2 =C_D[1] .Y_3 + C_D[\perp].Y_4\\
X_4 =C_D[\perp].X_2 & Y_4 =C_D[\perp].Y_2\\
X_3 =GEN[M,N].X_5 & Y_3 =GEN[M,N].Y_5\\
X_5=C_D[M].X_6+C_D[\perp].X_7 & Y_5=C_D[M].Y_6+C_D[\perp].Y_7\\
X_7 =(C_D[0]+c_D[\perp]) .X_3 & Y_7 =(C_D[1]+c_D[\perp]) .Y_3\\
X_6 =Me[A,q_i; kl].X_8 & Y_6 =Me[A,q_i; kl].Y_8\\
X_8 =\sigma_{kl}[M] .X_{9} & Y_8 =\sigma_{kl}[M] .Y_{9}\\
X_9=C_D[M].X_{10}+C_D[\perp].X_{11}  & Y_9=C_D[M].Y_{10}+C_D[\perp].Y_{11} \\
X_{11} =GEN[M,N].X_{12} & Y_{11} =GEN[M,N].Y_{12}\\
X_{12}=C_D[M].X_{13}+C_D[\perp].X_7  & Y_{12}=C_D[M].Y_{13}+C_D[\perp].Y_7 \\
X_{13}=\sigma_{kl}[M].X_9  & Y_{13}=\sigma_{kl}[M].Y_9 \\
X_{10}=C_D[0].X_{14}+C_D[\perp] .X_{15} & Y_{10}=C_D[1].Y_{14}+C_D[\perp] .Y_{15}\\
X_{15} =Me[M,N; kl].X_{16} & Y_{15} =Me[M,N; kl].Y_{16}\\
X_{16}=\sigma_{kl}[B].X_{17} & Y_{16}=\sigma_{kl}[B].Y_{17}\\
X_{17} =send_P[B].X_{18} & Y_{17} =send_P[B].Y_{18}\\
X_{18} =(C_D[0]+C_D[\perp]).X_4 & Y_{18} =(C_D[1]+C_D[\perp]).Y_4\\
X_{14}=Me[M,N;kl].X_{19} & Y_{14}=Me[M,N;kl].Y_{19}\\
X_{19} =\sigma_{kl}[B].X_{20} & Y_{19} =\sigma_{kl}[B].Y_{20}\\
X_{20} =send_P[B].Y_1 & Y_{20} =send_P[B].X_1\\
 & \\
\hline
\end{array}$$
\begin{remark}
As a result of quantum teleportation, the particle B in the last line is the same $q_i$.
\end{remark}

We proceed to prove that the process term $\tau_I(\langle X_1|E \rangle)$ exhibits the desired
external behaviour of the protocol. After applying the abstraction operator
$\tau_I$ to the process term $\langle X_1|E \rangle$, the loops of communication actions become $\tau-$loops. These loops can be removed using CFAR. 
\\\\
$\tau _I(\langle X_1|E\rangle)=\sum\limits_{q_i \in \Delta } read_Q[q_i].\tau _I(\langle X_2|E\rangle)
\\\stackrel{CFAR}{=}\sum\limits_{q_i \in \Delta } read_Q[q_i]. \tau _I(\langle X_3|E\rangle)$\\
$~~~=~~\sum\limits_{q_i \in \Delta } read_Q[q_i].\tau _I(\langle X_5|E\rangle)\\
\stackrel{CFAR}{=}\sum\limits_{q_i \in \Delta } read_Q[q_i].\tau _I(\langle X_6|E\rangle)$\\
$~~~=~~\sum\limits_{q_i \in \Delta } read_Q[q_i].\tau _I(\langle X_8|E\rangle)$\\
$~~~=~~\sum\limits_{q_i \in \Delta } read_Q[q_i].\tau _I(\langle X_9|E\rangle)\\
\stackrel{CFAR}{=}\sum\limits_{q_i \in \Delta } read_Q[q_i].\tau _I(\langle X_{14}|E\rangle)$\\
$~~~=~~\sum\limits_{q_i \in \Delta } read_Q[q_i].\tau _I(\langle X_{19}|E\rangle)\\
~~~=~~\sum\limits_{q_i \in \Delta } read_Q[q_i].\tau _I(\langle X_{20}|E\rangle)\\
~~~=~~\sum\limits_{q_i \in \Delta } read_Q[q_i].send_P[q_i].\tau _I(\langle Y_1|E\rangle)$\\

Likewise, applying RDP, QTI1-QTI5, the following equation can be derived:
$$\tau _I(\langle Y_1|E\rangle)=\sum\limits_{q_i \in \Delta } read_Q[q_i].send_P[q_i].\tau _I(\langle X_1|E\rangle)$$
From the last two equations together with RSP, the following equation can be derived:
$$\tau _I(\langle X_1|E\rangle)=\sum\limits_{q_i \in \Delta } read_Q[q_i].send_P[q_i].\tau _I(\langle X_1|E\rangle)$$
In combination with equation (*) this yields
$$\tau _I(\partial_H(S(0)||R(0)))=\sum\limits_{q_i \in \Delta } read_Q[q_i].send_P[q_i].\tau _I(\partial_H(S(0)||R(0)))$$
Therefore, the QAQP exhibits the desired external behaviour. So, the verification of the QAQP is finished. 


\section{Acknowledgements}
Thanks in advance to respectful referees.


%
%
%
%
%
%
%

\begin{thebibliography}{}


%
%
%
%
%
%
%
%
%
%

\bibitem{Alternatingbit}
J.W. de Bakker and J.I. Zucker (1982), \textit{Processes and the denotational semantics of concurrency}, Information and Control, 54, pp. 70-120. 

\bibitem{Feng2014}
Y. Feng, Y. Deng and M. Ying (2014),\textit{ Symbolic Bisimulation for Quantum Processes}, ACM Transactions on Computational Logic, Vol. 15 (2), Article No. 14.
\bibitem{Fokkink}
W. Fokkink (2007), \textit{Introduction to process algebra}, Springer-Verlag.

\bibitem{Nielsen}
M. A. Nielsen and I. L. Chuang (2010), \textit{Quantum Computation and Quantum Information}, Cambridge university press (New York).

\bibitem{Wang2013}
Y. Wang (2013), \textit{An Axiomatization for Quantum Processes to Unifying Quantum and Classical Computing},
Manuscript, http://arxiv.org/abs/1311.2960. 

\bibitem{Wang2015}
Y. Wang (2015), \textit{Entanglement in Quantum Process Algebra}, Manuscript, http://arxiv.org/abs/1404.0665. 

\bibitem{Wilde2013}
M. M. Wilde (2013), Quantum Information Theory, Cambridge university press (New York).



\end{thebibliography}
\end{document}